\documentclass[12pt,epsf,amssymb]{article}
\usepackage{epsfig,psfrag}
\usepackage{graphicx}
\usepackage{amsfonts}
\usepackage{amsmath}
\usepackage[usenames]{color}
\usepackage{amssymb}

\begin{document}
\baselineskip 0.6cm
\newcommand{\gsim}{ \mathop{}_{\textstyle \sim}^{\textstyle >} }
\newcommand{\lsim}{ \mathop{}_{\textstyle \sim}^{\textstyle 3<} }
\newcommand{\vev}[1]{ \left\langle {#1} \right\rangle }
\newcommand{\bra}[1]{ \langle {#1} | }
\newcommand{\ket}[1]{ | {#1} \rangle }
\newcommand{\Dsl}{\mbox{\ooalign{\hfil/\hfil\crcr$D$}}}
\newcommand{\nequiv}{\mbox{\ooalign{\hfil/\hfil\crcr$\equiv$}}}
\newcommand{\nsupset}{\mbox{\ooalign{\hfil/\hfil\crcr$\supset$}}}
\newcommand{\nni}{\mbox{\ooalign{\hfil/\hfil\crcr$\ni$}}}
\newcommand{\EV}{ {\rm eV} }
\newcommand{\KEV}{ {\rm keV} }
\newcommand{\MEV}{ {\rm MeV} }
\newcommand{\GEV}{ {\rm GeV} }
\newcommand{\TEV}{ {\rm TeV} }

\newcommand{\be}[1]{\begin{equation}\label{#1}}
\newcommand{\ee}{\end{equation}}
\newcommand{\ds}{\displaystyle}
\newcommand{\bea}[1]{\begin{eqnarray}\label{#1}}
\newcommand{\eea}{\end{eqnarray}}
\newcommand{\ra}{\rightarrow}

\def\diag{\mathop{\rm diag}\nolimits}
\def\tr{\mathop{\rm tr}}

\def\Spin{\mathop{\rm Spin}}
\def\SO{\mathop{\rm SO}}
\def\O{\mathop{\rm O}}
\def\SU{\mathop{\rm SU}}
\def\U{\mathop{\rm U}}
\def\Sp{\mathop{\rm Sp}}
\def\SL{\mathop{\rm SL}}
\def\Qt{{\tilde Q}}
\def\mt{\tilde{m}}
\def\change#1#2{{\color{blue}#1}{\color{red} #2}\color{black}\hbox{}}


\begin{titlepage}
  
\begin{flushright}
LTH/981
\end{flushright}

\vskip 2cm
\begin{center}
{\large \bf  Flows between Dualities for 3d Chern-Simons Theories}
\vskip 1.2cm
Siraj Khan and Radu Tatar 

\vskip 0.4cm

{\it Division of Theoretical Physics, Department of Mathematical Sciences

The University of Liverpool,
Liverpool,~L69 3BX, England, U.K.

rtatar@liverpool.ac.uk, srk@liverpool.ac.uk}

\vskip 1.5cm

\abstract{We study Seiberg-like dualities for 3d  ${\cal{N}}=2$ theories with flavors in fundamental and adjoint representations. 
The recent results of Intriligator and Seiberg provide a derivation of an Aharony duality from a Giveon-Kutasov duality. 
We extend their result to the case of more general theories involving various masses for fundamental quarks and adjoint fields. By fine tuning the vev of a scalar field and using various identifications between gauge groups  and their singlet duals, we derive several
examples of Aharony dualities. 
For theories with an adjoint field, we discuss the connection between  the Aharony dualities proposed by Kim and Park  for theories with  multiple Coulomb branches and Giveon-Kutasov-Niarchos dualities.
} 

\end{center}
\end{titlepage}


\section{Introduction}

The breaking of SUSY from ${\cal{N}} = 2$ to ${\cal{N}} = 1$  in four dimensional theories with flavors has been the subject of intense study in 
both field  theory or string theory. Is is achieved by turning on superpotentials for the adjoint fields and the matter fields. 
In string theory, one can either consider brane configurations with suspended D4 branes \cite{giku},
geometries with wrapped D5 branes \cite{vafa} or matrix models \cite{dv}. 
These distinct approaches are related to each other  by T-dualities \cite{rt}. String theory provides insights 
into strongly coupling regimes, Seiberg dualities, metastable vacua, etc. 

One immediate question is whether one can translate this knowledge to 3 dimensional theories, where the corresponding deformation would be  
${\cal{N}} = 4 \rightarrow {\cal{N}} = 2$. A partial answer to this question was obtained some time ago when certain aspects of mirror symmetry 
for 3d ${\cal{N}} = 2$ theories were uncovered \cite{is1}.
In recent years there has been a renewal of interest in 3d gauge theories along various exciting research avenues. 
On one hand 3-algebras were shown to describe 
multiple, parallel membranes in M-theory (see \cite{lmc} for a review and a complete list of references). 
On the other hand, 
a new AdS/CFT duality was proposed in \cite{abjm} (see again \cite{lmc} for details and references).
 
A third recent direction, built up on the developments of \cite{is1}, has already provided important fresh tools for exploring 
Seiberg dualities as the computation of the partition functions on 3-spheres 
and the corresponding superconformal index \cite{kap}-\cite{jaffe}. 
Various connections between  3d dualities and usual 4d dualities were discussed in \cite{3d4d,3d4d1}. 
    
In 3-dimensions there are two types of Seiberg-like dualities:

$\bullet$ Giveon-Kutasov duality which applies for theories with  any Chern-Simons level. This is reminiscent of a 4d Seiberg duality 
for theories with fundamental matter \cite{gk1} or an adjoint field \cite{niarchos}. The Giveon-Kutasov duality  
is well understood from brane configurations with D3 branes suspended between NS branes.

$\bullet$ Aharony duality for 3d  which has a peculiar coupling between electric and magnetic monopoles \cite{aha1}. This does not have a clear brane picture.  

The two Seiberg-like dualities can be connected by starting with the Aharony duality, adding masses and generating Chern-Simons terms, resulting in the Giveon-Kutasov duality. It is less clear how to proceed the other way 
because a coupling between electric and magnetic monopoles has to be generated. Very recently, \cite{is} discussed a reverse RG flow from Giveon-Kutasov duality in the UV to Aharony duality in the IR. 
This flow provided a derivation of the monopole operator in Aharony duality. They considered the case of 
one massive fundamental flavor and performed a fine  tuning of the expectation value for the extra adjoint 
scalar coming from the 4d ${\cal{N}} = 2$ vector multiplet. The result was an extra  set of $U(1)$ groups 
with Chern-Simons level $k= -\frac{1}{2}$ which are dual to some singlets \cite{guk1,benini1}, 
subsequently identified with electric monopoles in the magnetic theory. 

Our work extends the discussion of \cite{is} in two directions. Firstly we consider the case when more than one flavor is massive (we consider the case of 
two and four massive flavors). The choices of mass and the tuning of the corresponding vev for 
the adjoint scalar field lead to various electric  
theories and their corresponding Giveon-Kutasov duals. For particular values of the Chern-Simons level, some unbroken groups are of type $U(k)_{-\frac{k}{2}}$ and have dual descriptions in terms of singlets \cite{benini1} which are identified 
with the Aharony dual monopoles. 
The second direction that we consider involves theories with an extra adjoint field where the Aharony duality has multi-dimensional 
Coulomb branches and multiple monopole operators are needed \cite{park1}. The Giveon-Kutasov duality for Chern-Simons theories with adjoint matter 
was proposed in \cite{niarchos}. 
We explain how masses for fundamental fields provide unbroken groups with singlet duals 
providing pairs of monopoles and connecting the two duality pairs. We use the continuous deformation between the 
adjoint field potential $\mbox{Tr}~\Phi^n$ and the one with lower powers in $\Phi$ in order to 
to understand how pairs of monopoles overlap. 

\section{Field Theory Results}

We start with ${\cal{N}} = 4$ theory in 3 dimensions obtained by suspending $N$ D3 branes between  
two parallel  NS branes with extra $N_f$ flavor D5 branes. 

In ${\cal{N}} = 2$ language, the hypermultiplets can be written as two chiral superfields $Q$ and $\Qt$
with charges $1$ and $-1$.  The ${\cal{N}} = 4$ vector multiplet contains an ${\cal{N}} = 2$
real vector multiplet $V$ with a real scalar $\sigma$ as the lowest component and
a chiral multiplet $\Phi$ with a complex scalar as the lowest component. 
Another type of ${\cal{N}} = 4$ multiplet is the linear multiplet which contains 
an ${\cal{N}} = 2$ linear multiplet $\Sigma$ (which is  real and satisfies
$D^2\Sigma=\bar D^2\Sigma=0$).

The kinetic part of the Lagrangian for an ${\cal{N}} = 4$ theory is
\be{kinetic}
L_{kin} = \int d^4\theta 
\left(Q^\dag e^{2V} Q + \Qt^\dag e^{-2V}\Qt\right) 
+ \left[\int  d^2\theta {\sqrt 2}\Phi Q\Qt + 
c.c.\right].\ 
\ee
The hypermultiplets  can have a complex mass $m_{c}$ coming form their coupling to the multiplet $\Phi$ or a 
real mass $m_r$:
\be{mass}
L_{mass} = \int d^4\theta
\left(Q^\dag e^{ 2im_r \theta \bar{\theta}} Q + 
\Qt^\dag e^{- 2im_r \theta\bar{\theta}}\Qt\right)
+ \left[\int d^2\theta m_{c} Q\Qt +c.c.\right].
\ee

There are also Fayet-Iliopoulos terms
\be{fi4}
\xi_{V} \int d^4 \theta~V + \xi_{\Phi} \int d^2 \theta~\Phi
\ee
and Chern-Simons terms  
\be{cs}
\int d^4 \theta \frac{k}{4 \pi} \Sigma~V
\ee

The ${\cal{N}} = 4$ theory is deformed to an ${\cal{N}} = 2$ one by adding a mass term for the field $\Phi$ 
\footnote{We will consider more general deformations in a later section.} 
\be{deff}
\int d^2 \theta \frac{1}{2} \mu^2 \Phi^2.
\ee
Integrating out the massive fields, we obtain ${\cal{N}} = 2, U(N)$ theory with $N_f$ flavor fields. 

The terms containing $\Phi$ are
\be{phi}
\int d^2 \theta ({\sqrt 2}\Phi Q\Qt + \frac{1}{2} \mu^2 \Phi^2)
\ee
The equation of motion for $\Phi$ is
\be{phi11}
{\sqrt 2} Q\Qt + \mu \Phi = 0
\ee 
with the simple solution $Q\Qt = \Phi = 0$, suggesting either  $Q = \Phi = 0$ or  $\Qt = \Phi = 0$.

In terms of brane configurations, we consider that one NS brane rotates by an angle 
$\tan(\theta) = \mu$ which goes to $\infty$ for ${\cal{N}} = 2$ theory. In this limit, the 
equation of motion for the remaining fields in the  ${\cal{N}} = 2$ theory are \cite{is}:
\be{iseqt}
\int d^4 \theta (V + m_r) Q = 0;~~~~~\int d^4 \theta (- V + m_r) \tilde{Q} = 0   
\ee
which for the scalar components of $Q$ and $\tilde{Q}$ are \cite{is}
\be{lowest}
(\sigma + m_r) Q = 0;~~~~~(- \sigma + m_r) \tilde{Q} = 0
\ee
We have the following possibilities:

1) $\sigma = 0, Q = 0, \tilde{Q} = 0$,

2)~$Q = 0,  \sigma =  m_r$,

3)~$\tilde{Q} = 0, \sigma = - m_r$.  

\vskip .1cm

We also impose the D-term equation which becomes

\be{dterm}
- \frac{k}{2\pi} \sigma -  \frac{\xi}{2\pi} + Q^\dag Q - \Qt^\dag\Qt = 0
\ee

\section{Review of Intriligator-Seiberg Result }

We are reviewing the Seiberg duality flow discussion for one massive flavor described in \cite{is}.

Consider first the case when only one flavor gets massive and we have zero $\sigma, Q^{N_f}, \tilde{Q}^{N_f}$. In brane configurations we have a 
stack of $N$ D3 branes at $\sigma = 0$. 
On the other hand, integrating out the  massive flavor causes  the Chern-Simons level to increase by 1.  The flavor D5 branes split 
into (1,1) branes \cite{bh} and the gauge group has a bigger Chern-Simons level, becoming $U(N)_1$. If the initial level was $k$, a $(1,k)$ brane splits into $(1,k+1)$ branes and the gauge group becomes  $U(N)_{k+1}$.

When $\sigma = \pm m_r$, some of the D3 branes move to the points $\pm m_r$. For one massive flavor, the gauge group is broken to $U(N_c - 1)_{k+1} \times U(1)_{k+\frac{1}{2}}$. $U(N_c - 1)$ group has
$N_f - 1$ massless flavors and one massive flavor $Q^{N_f}, \tilde{Q}^{N_f}$. $U(1)$ only has either $Q^{N_f}$ or
$\tilde{Q}^{N_f}$ as massive flavors. After integrating out the massive flavor, the level of the  $U(N_c - 1)$ gauge group increases by 1 whereas 
the level of the $U(1)$ gauge group changes by  $+\frac{1}{2}$. In brane configuration language, the  $U(N_c - 1)$ group lives on the 
 $N_c - 1$ D3 branes which are not moved whereas $U(1)$ lives on the single D3 brane moved and suspended between an NS brane and the intersection
 point of $(1,k)$ 5 brane, D5 brane and $(1,k+1)$ 5 branes.  

For $\sigma = \pm m_r$  we also need to consider the D-term equation (\ref{dterm}).   
When $Q^\dag Q$ and $\Qt^\dag\Qt$ are zero, the value for the Fayet-Iliopoulos parameter is $\xi = m$. 
In brane configurations, the Fayet-Iliopoulos parameter
is the distance between the two NS branes. It is generally associated to a supersymmetry breaking. 
Nevertheless, in our present case the Fayet-Iliopoulos term balances the non-zero expectation value for the field 
$\sigma$ and does not break supersymmetry.

The $U(N_c - 1)_{k+1} \times U(1)_{k+\frac{1}{2}}$ brane configuration contains:

- one $(1, k+ 1)$ 5-brane between two junctions of D5 intersecting $(1, k)$ branes,

- three rotated NS branes: NS$_0$ and NS$_{\pm}$,

- $N_c - 1$ D3 branes between NS and NS$_{0}$,

- one D3 brane between NS and either NS$_{+}$ or NS$_{-}$.

\subsection{Giveon-Kutasov to Aharony Duality for One Massive Flavor}
 
\cite{is} considered  the flow between Giveon-Kutasov \cite{gk1} and Aharony dualities
\cite{aha1}. 

The Giveon-Kutasov duality maps ${\cal{N}} = 2$, $U(N_c)$ Chern-Simons theory with $N_f$ flavors and level $k$  
to a  dual theory $U(N_f + |k| - N_c)_{-k}$ with a superpotential $M q \tilde{q}$, where $q$ are the 
dual quarks and $M$ is a singlet.  On the other hand, the Aharony duality maps   ${\cal{N}} = 2$, $U(N_c)$ theory with $N_f$ flavors and level 0 to  $U(N_f-N_c)$ with an unusual superpotential coupling the electric $X_{\pm}$ and magnetic monopoles $\tilde{X}_{\pm}$:
\be{ahapot}
M q \tilde{q} + X_+ \tilde{X}_- + X_- \tilde{X}_+ 
\ee
One can flow from Aharony duality to Giveon-Kutasov duality by turning on real masses for the electric 
flavors. It is less clear how to flow from  Giveon-Kutasov duality to Aharony duality and the solution was offered in \cite{is}:

$\bullet$ Start with Giveon-Kutasov electric theory $U(N_c)_k$ with $N_f$ quark flavors and add a real mass for one of the 
flavor $Q^{N_f}$ and $\tilde{Q}_{N_f}$:

$\bullet$ For $\sigma = 0, Q = 0, \tilde{Q} = 0$, the Giveon-Kutasov electric theory modifies into an Aharony electric theory 
$U(N_c)_{k+1}$ with $N_f-1$  flavors, when $k=-1$.

$\bullet$ For  $\sigma = \pm m_r$, the Giveon-Kutasov magnetic theory becomes, 
\be{iss}
U(N_f-N_c -1)_{k+1} \times U(1)_{k+\frac{1}{2}}  \times U(1)_{k+\frac{1}{2}}
\ee
This is to be identified with the Aharony magnetic theory when $k=-1$. The identification holds for $k=-1$ and the proposed Aharony duality takes 
$U(N_c)_{0}$ with $N_f-1$  flavors into  $U(N_f-N_c -1)_{0} \times U(1)_{-\frac{1}{2}}  \times U(1)_{-\frac{1}{2}}$ with $N_f - 1$ flavors. 
The product $U(1)_{-\frac{1}{2}}  \times U(1)_{-\frac{1}{2}}$ is dual to a pair of single chiral superfields $X_{\pm}$  \cite{guk1,benini1}. 
The single superfields couple to the magnetic monopoles $\tilde{X}_{\pm}$ associated to 
$U(1)$ inside $U(N_c - 1)$ as $X_{-} \tilde{X}_++ X_{+} \tilde{X}_-$, as required for Aharony duality.

\section{Flow between Dualities for Two Massive Flavors}

We now generalize the results of \cite{is} to the case of more massive flavors. We start with two massive flavors  
$Q^{N_f}, Q^{N_f -1}$ and $\tilde{Q}_{N_f},\tilde{Q}_{N_f-1}$ with masses 
\be{2flavors}
m_{1~f}^{f'} = m_1 \delta_{f N_f}^{f' N_f};~~m_{2~f}^{f'} = m_2 \delta_{f N_{f}-1}^{f' N_{f}-1},~~
\mt_{1~f}^{f'} = m_1 \delta_{\tilde{f} N_f}^{\tilde{f}' N_f};~~\mt_{2~f}^{f'} = m_2 \delta_{\tilde{f} N_{f}-1}^{\tilde{f}' N_{f}-1}  
\ee   
We start with the case of equal masses $m_1 = m_2 = m$. 

\subsection{Equal Masses $m_1=m_2$}

If the masses are equal $m_1 = m_2 = m$, we can have one of the following (all entries of $Q, \tilde{Q}$ are always zero):

Case 1) All components of $\sigma$ are 0. 

After integrating out the massive flavors the gauge group becomes $U(N_c)_{k+2}$. This group lives on $N_c$ D3 branes 
suspended between an NS brane and a $(1,k+2)$ 5-brane. 

Case 2) $\sigma^{N_c}_{N_c}= \sigma^{N_c-1}_{N_c-1}$ are $- m$. The other components are 0.  

The gauge group becomes $U(N_c - 2)_{k+2} \times U(2)_{k+1}$ group. The unbroken $U(2)_{k+1}$ is a decoupled factor with two $\tilde{Q}$ fields with mass
$2 m$  and two $Q$ fields with mass 0. By integrating out the massive flavors $\tilde{Q}$, we get a modification in the level and 
Fayet-Iliopoulos 
term:
\be{modif2}
k \rightarrow k+1,~~\xi \rightarrow \xi + 2 m
\ee
which, after a shift, relates the FI term and the mass $m$ as $\xi = m(k+2)$.  

We analyze the low energy $U(2)_{k+1}$ theory. The Witten index for the theory with two massless $Q$ fields and level $k+1$
is  \cite{is}
\be{wit}
\mbox{Tr}~(-1)^F = |k+1| +1
\ee
which is non-zero for any value of $k$. The theory has SUSY vacua but they can be topological. For $|k+1|=1$, if one adds
massive matter and flows down, the theory at the origin is an IR-free theory consisting of one chiral superfield with 
no superpotential. This superfield is to be  identified with the electric monopole $X_-$. 
 
One special case is $k=-2$, when the broken gauge group becomes  $U(N_c - 2)_{0} \times U(2)_{-1}$.
As in \cite{benini1}, $U(2)_{-1}$ with two $Q$ fields has a magnetic dual with only one singlet $X_-$ which couples to the 
operator $\tilde{X}_+$ associated to the $U(1)$ inside $U(N_c -2)$. $X_-$ is exactly the chiral superfield describing the 
vacuum of $U(2)_{-1}$. 
 
The Chern-Simons level of the field theory on the D3 branes is: 

- $k$ for D3 branes  between the NS brane and the $(1,k)$ brane,

- $k+2$ for D3 branes between the NS brane and the $(1,k+2)$ brane,

- $k+1$ for D3 branes between the NS brane and the junction point of $(1,k), (1,k+2)$ and 2 D5 branes

Case 3) $\sigma^{N_c}_{N_c}= \sigma^{N_c-1}_{N_c-1} = + m$ and  $\sigma^{N_c-2}_{N_c-2}= \sigma^{N_c-3}_{N_c-3} = - m$. 
The gauge group becomes $U(N_c - 4)_{k+2} \times U(2)_{k+1} \times U(2)_{k+1}$.  For the case $k = -2$, the duals for both $U(2)_{-1}$ provide two chiral superfields $X_-, X_+$.

Case 4) $\sigma^{N_c}_{N_c} = \pm m$ and  
$\sigma^{N_c-1}_{N_c-1} = 0$. The two massive flavors are integrated out and one obtains the 
product between $U(N_c-1)_{k+2}$ with $N_f - 2$ flavors and $U(1)_{k+1}$ with 2 massless $Q$ fields.
The Witten index for $U(1)_{k+1}$ is $|k+1| + 1$ which is always positive so the supersymmetry is preserved.  
Nevertheless, for  $k=-2$, the Abelian group becomes $U(1)_{-1}$ and no candidate for magnetic monopoles exists. 

Case 5)   $\sigma^{N_c}_{N_c} = \pm m,~\sigma^{N_c-1}_{N_c-1} = \mp m$ and  
$\sigma^{N_c-2}_{N_c-2} = 0$. After integrating out the two massive flavors, the remaining gauge group is 
 $U(N_c-2)_{k+2} \times U(1)_{k+1} \times U(1)_{k+1}$

 \subsection{Different Masses for the Two Flavors} 

We now consider the case when the mass for $Q^{N_f}$ flavor is bigger than the one for $Q^{N_f - 1}$. 

Case 1) remains the same as for equal masses.

For Case 2)  we choose two nonzero entries for $\sigma$ as $\sigma^{N_c}_{N_c}= - m_1, \sigma^{N_c-1}_{N_c-1} = -  m_2$.
The unbroken gauge group is $U(N_c - 2)_{k+2} \times U(1)_{k_1} \times U(1)_{k_2}$ where $k_{1,2}$ need to be determined.

The  $U(1)_{k_1}$ on the D3 brane displaced to  $\sigma^{N_c}_{N_c}= - m_1$ has two massive $\tilde{Q}$ fields with positive masses $2 m_1$ 
and $m_1 + m_2$ respectively. The effective real masses for the fields $Q$ are
\be{effective}
m_{i}(\sigma) = m_i + n_i \sigma, \mbox{where} n_i~~\mbox{is the charge under}~~U(1)
\ee
so the effective mass for  $Q^{N_f}$ is 0 and the one for  $Q^{N_f-1}$ is $m_2 - m_1 < 0$.  The  $U(1)_{k_2}$ on the D3 brane displaced to  
$\sigma^{N_c-1}_{N_c-1}= - m_2$ contains two massive $\tilde{Q}$ fields with positive masses $2 m_2$ and $m_1 + m_2$ respectively and one field $Q$ with positive mass. 
The effective mass for the fields $Q$ is  $m_1 - m_2 > 0$ for  $Q^{N_f}$ and zero for  $Q^{N_f-1}$.
The effective Chern-Simons level 
\be{cslevel}
k_{eff} = k + \frac{1}{2} \sum_i n_i^2~\mbox{sign}~(m_i(\sigma))  
\ee
implies that $k_1 = k + \frac{1}{2}, k_2 = k + \frac{3}{2}$ and the unbroken gauge group becomes  
\be{dimasses}
U(N_c - 2)_{k+2} \times U(1)_{k+\frac{1}{2}} \times U(1)_{k+\frac{3}{2}}
\ee

To connect to Aharony duality, we consider $k = - 2$ and the unbroken gauge group becomes
$U(N_c - 2)_{0} \times U(1)_{-\frac{3}{2}} \times U(1)_{-\frac{1}{2}}$. $U(1)_{-\frac{3}{2}}$ and $U(1)_{-\frac{1}{2}}$ have two $\tilde{Q}$ 
fields and one $Q$ field. $U(1)_{-\frac{1}{2}}$ has a dual with a free singlet field $X_+$. What about the group  
$U(1)_{-\frac{3}{2}}$? The effective FI term is $k_{eff} \sigma + \xi_{eff}$ which is negative so would signal no SUSY vacuum.    
Nevertheless, as   $U(1)_{-\frac{3}{2}}$ does not contribute to the Aharony duality, it can decouple and the theory would remain SUSY. It would be interesting to discuss this decoupling in detail.

Case 3)   All entries of $Q$ are 0 and two nonzero entries of $\sigma$ are $\sigma^{N_c}_{N_c}= + m_1, \sigma^{N_c-1}_{N_c-1} = +  m_2$.
The unbroken gauge group is $U(N_c - 2)_{k+2} \times U(1)_{k_3} \times U(1)_{k_4}$.

The  $U(1)_{k_3}$ on D3 branes displaced to  $\sigma^{N_c}_{N_c}= m_1$ contains two massive $Q$ fields with positive masses $2 m_1$ and $m_1 + m_2$ respectively 
together with a $\tilde{Q}$ field with negative mass  $m_2 - m_1$. The value for $k_3$ is $k+\frac{1}{2}$. The $U(1)_{k_4}$ on D3 branes located at
  $\sigma^{N_c}_{N_c}= m_2$ has two massive $Q$ fields with positive mass $2 m_2$ and $m_1 + m_2$ respectively plus a $\tilde{Q}$ field with positive mass  
$m_1 - m_2$. The value for $k_3$ is then $k+\frac{3}{2}$ so the unbroken gauge group 
$U(N_c - 2)_{k+2} \times U(1)_{k+\frac{1}{2}} \times U(1)_{k+\frac{3}{2}}$. For $k=-2$, the group becomes 
$U(N_c - 2)_{0} \times U(1)_{-\frac{3}{2}} \times U(1)_{-\frac{1}{2}}$.

Case 4) The nonzero entries of $\sigma$ are  
$\sigma^{N_c}_{N_c}= m_1, \sigma^{N_c-1}_{N_c-1} = m_2, \sigma^{N_c-2}_{N_c-2}= - m_2, \sigma^{N_c-3}_{N_c-3} = - m_1$. The unbroken gauge group is 
 $U(N_c - 2)_{k+2} \times U(1)_{k_5} \times U(1)_{k_6}  \times U(1)_{k_7} \times U(1)_{k_8}$. The field content for all $U(1)$ groups imply that 
\be{kval}
k_5 = k_8 = k + \frac{1}{2};~~~k_6 = k_7 = k + \frac{3}{2}  
\ee
For $k=-2$, the unbroken group becomes
\be{ung}
 U(N_c - 4)_{0} \times U(1)_{-\frac{3}{2}} \times U(1)_{-\frac{1}{2}}  \times U(1)_{- \frac{1}{2}} \times U(1)_{- \frac{3}{2}}
\ee
The two  $U(1)_{-\frac{1}{2}}$ groups have duals represented by the scalars $X_{\pm}$. 

Case 5) Other interesting cases are 
$\sigma^{N_c}_{N_c}= 0, \sigma^{N_c-1}_{N_c-1} = - m_1$ or $\sigma^{N_c}_{N_c}= 0, \sigma^{N_c-1}_{N_c-1} = - m_2$ in which case 
the unbroken gauge group is $U(N_c-1)_{k+2} \times U(1)_{k+\frac{1}{2}}$ or  $U(N_c-1)_{k+2} \times U(1)_{k+\frac{3}{2}}$. 
For $k=-2$, this becomes $U(N_c - 1)_{0} \times U(1)_{-\frac{3}{2}}$ or $U(N_c - 1)_0 \times U(1)_{-\frac{1}{2}}$.

For all cases 3),4) and 5) we need to decouple the $U(1)_{-\frac{3}{2}}$ in order to save the SUSY. 

\subsection{Giveon-Kutasov to Aharony Duality for 2 Massive Flavors}

We start with Giveon-Kutasov electric theory with gauge group $U(N_c)_{-2}$ with $N_f$ quark flavors. The $N_f$ and 
$N_f-1$ flavors have a real mass. After integrating out the massive flavors and considering  a
vacuum with $\sigma = 0$, the electric theory becomes an Aharony electric theory with $U(N_c)_{0}$ with $N_f - 2$ massless flavors. 

One the other hand, the Giveon-Kutasov magnetic dual is $U(N_f + 2 - N_c)_{-2}$. How can we reach the Aharony magnetic dual?
The answer depends on the choice for the masses:

$\bullet$ If the two masses are equal, the corresponding case 3 provides an Aharony magnetic dual  with a gauge group     
$U(N_f - 2-N_c)_{0} \times U(2)_{-1} \times U(2)_{-1}$ with $N_f - 2$ flavors. Each $U(2)_{-1}$ is dual to a singlet $X_{\pm}$ which contributes to 
the mysterious superpotential. Therefore, the corresponding case 1) for electric theory with equal masses 
is dual to case 3) for magnetic theory with equal masses. This is the Aharony duality for equal flavor masses. 
 
$\bullet$ If the two masses are equal and we consider case 4), the electric  theory would be $U(N_c-1)_{0} \times U(1)_{-1}$ and the magnetic one
$U(N_f + 1 - N_c)_0 \times U(1)_{-1}$. This now looks like an Aharony duality. 

$\bullet$ If the masses are not equal, we consider the case 3) for non equal masses and obtain a theory with a gauge group 
(\ref{ung}). Both  $U(1)_{-\frac{1}{2}}$ groups can be dualised into singlets and we obtain $U(N_f - 2 - N_c)$ with singlets and superpotential, 
together with an extra 
product of two $U(1)_{- \frac{3}{2}}$. The Aharony duality is valid only if the two masses are very large and almost equal. The 
product of two $U(1)_{- \frac{3}{2}}$ can decouple to preserve SUSY. 

$\bullet$ If we are in the case 5) of non equal masses, the electric theory is  $U(N_c - 1)_{0} \times U(1)_{-\frac{3}{2}}$ or 
$U(N_c - 1)_0 \times U(1)_{-\frac{1}{2}}$ and the magnetic theory is  $U(N_f+1-N_c)_{0} \times U(1)_{-\frac{3}{2}}$ or 
$U(N_f + 1 - N_c)_0 \times U(1)_{-\frac{1}{2}}$.  We have Giveon-Kutasov duality between $U(N_c - 1)_0 \times U(1)_{-\frac{1}{2}}$ and 
$U(N_f + 1 - N_c)_0 \times U(1)_{-\frac{1}{2}}$ if both $U(1)_{-\frac{1}{2}}$ are dualised into singlets. For $U(N_c - 1)_{0} \times U(1)_{-\frac{3}{2}}$ and
 $U(N_f+1-N_c)_{0} \times U(1)_{-\frac{3}{2}}$, we have a Giveon-Kutasov duality if the $U(1)_{-\frac{3}{2}}$ groups decouple to save SUSY. 

The conclusion is that by starting from a Giveon-Kutasov duality we can reach a multitude of dualities, depending on the choice of the vevs for
 $\sigma$. 

\section{Flow between Dualities for Four Massive Flavors}

We now move to a more involved case with four massive flavors $Q^{N_f}, Q^{N_f -1}, Q^{N_f -2}, Q^{N_f-3} $ and 
$\tilde{Q}_{N_f},\tilde{Q}_{N_f-1}, \tilde{Q}^{N_f -2}, \tilde{Q}^{N_f -3} $ as
\be{4flavours}
m_{1~f}^{f'} = m_1 \delta_{f N_f}^{f' N_f},~~m_{2~f}^{f'} = m_2 \delta_{f N_{f}-1}^{f' N_{f}-1},~~m_{3~f}^{f'} = m_3 \delta_{f N_{f}-2}^{f' N_{f}-2}, ~m_{4~f}^{f'} = m_4 \delta_{f N_{f-3}}^{f' N_{f}-3}
\ee
\be{3flavourst}
\mt_{1~f}^{f'} = m_1 \delta_{\tilde{f} N_f}^{\tilde{f}' N_f},~~\mt_{2~f}^{f'} = m_2 \delta_{\tilde{f} N_{f}-1}^{\tilde{f}' N_{f}-1},~~
\mt_{3~f}^{f'} = m_3 \delta_{\tilde{f} N_f-2}^{\tilde{f}' N_f-2},~~\mt_{4~f}^{f'} = m_4 \delta_{f N_{f-3}}^{f' N_{f}-3} 
\ee   

We have several possibilities, among which we consider:

$\bullet$ Case (a): all the masses are equal $m_1 = m_2 = m_3 = m_4 = m$

$\bullet$ Case (b): three masses are equal. 

$\bullet$ Case (c): two masses are equal. 

$\bullet$ Case (d): all the masses are different and choose   $m_1 > m_2 > m_3 > m_4$.
 
\subsection{Case (a): all four masses are equal}

$\bullet$ 1) All components of $\sigma$ are 0, and all entries of $Q, \tilde{Q}$ are 0. 

After integrating out the massive flavors, the gauge group changes from $U(N_c)_k$ with $N_f$ flavors to  $U(N_c)_{k+4}$ with $N_f-4$ massless flavors. One special case is $k=-4$ where the level becomes 0. 

$\bullet$ 2) All entries of $Q$ are 0 and the $\sigma$  nonzero entries are
\be{4val}
\sigma^{N_c}_{N_c}= \sigma^{N_c-1}_{N_c-1} = \sigma^{N_c-2}_{N_c-2}= \sigma^{N_c-3}_{N_c-3} = - m.
\ee 
The gauge group is $U(N_c - 4)_{k+4} \times U(4)_{k+2}$. The unbroken $U(4)_{k+2}$ has four $\tilde{Q}$ fields with mass
$2 m$  and four $Q$ fields with mass 0 which implies an effective Chern-Simons level $k+2$ and an effective shifted FI term 
$\xi = m(k+4)$. 

One special case is $k=-4$ when the gauge group becomes  $U(N_c - 4)_{0} \times U(4)_{-2}$.
$U(4)_{-2}$ theory with four $Q$ fields has a dual with a singlet $X_-$. It would be very interesting to check 
that the Witten index is indeed non-zero for this case and we expect that the IR description is obtained by adding massive matter
and flowing down as in \cite{is} for $U(2)$ groups. 

A slight generalization is written as 
\be{2}
\sigma^{N_c}_{N_c}= \sigma^{N_c-1}_{N_c-1} =\sigma^{N_c-2}_{N_c-2}= \sigma^{N_c-3}_{N_c-3}= m,
\sigma^{N_c-4}_{N_c-4}= \sigma^{N_c-5}_{N_c-5} =\sigma^{N_c-6}_{N_c-6} = \sigma^{N_c-7}_{N_c-7}=- m,
\ee
when the gauge group becomes  $U(N_c - 8)_{k+4} \times U(4)_{k+2} \times U(4)_{k+2} $. For   $k=-4$ this is  
$U(N_c - 8)_{0} \times U(4)_{-2} \times U(4)_{-2}$ and both $U(4)_{-2}$ groups have duals with singlets $X_{\pm}$.

$\bullet$ 3) Some special choices for $\sigma^{N_c}_{N_c}, \sigma^{N_c-1}_{N_c-1}, \sigma^{N_c-2}_{N_c-2}, \sigma^{N_c-3}_{N_c-3}$. 
The unbroken gauge group is:

\vskip .1cm

- for $\sigma^{N_c}_{N_c} = - m, \sigma^{N_c-1}_{N_c-1} = \sigma^{N_c-2}_{N_c-2} = \sigma^{N_c-3}_{N_c-3}=0$:~~~$U(N_c - 1)_{k+4} \times U(1)_{k+2}$

\vskip .1cm

- for $\sigma^{N_c}_{N_c} = \sigma^{N_c-1}_{N_c-1} = - m, \sigma^{N_c-2}_{N_c-2} = \sigma^{N_c-3}_{N_c-3}=0$:~~~$U(N_c - 2)_{k+4} \times U(2)_{k+2}$

\vskip .1cm

- for $\sigma^{N_c}_{N_c} = \sigma^{N_c-1}_{N_c-1} = \sigma^{N_c-2}_{N_c-2} = - m, \sigma^{N_c-3}_{N_c-3}=0$:~~~$U(N_c - 3)_{k+4} \times U(3)_{k+2}$

The cases with eigenvalues equal to $m$ or mixed $m$ and $-m$ are similar.

\subsection{Case (b): Three equal masses}

$\bullet$ $m_1 = m_2 = m_3 > m_4$

Choose $\sigma^{N_c}_{N_c}= \sigma^{N_c-1}_{N_c-1} = \sigma^{N_c-2}_{N_c-2} = - m_1,  \sigma^{N_c-3}_{N_c-3} = - m_4$. In this case, for the first set 
of vevs of $\sigma$ one gets a $U(3)$ theory with 4 $\tilde{Q}$ fields with positive mass and one $Q$ field with negative mass which are integrated out.
The effective level is $k+\frac{3}{2}$. For $\sigma^{N_c-3}_{N_c-3} = - m_4$, we get $U(1)$ with 4  $\tilde{Q}$ with positive masses and three
 $Q$ with positive masses so the effective level is  $k+\frac{7}{2}$. The gauge group becomes 
\be{3masses}
U(N_c - 4)_{k+4} \times U(3)_{k+\frac{3}{2}} \times U(1)_{k+\frac{7}{2}}
\ee
For  $k = -4$ this becomes
 \be{3massesm41}
U(N_c - 4)_{0} \times U(3)_{- \frac{5}{2}} \times U(1)_{-\frac{1}{2}}
\ee
To preserve SUSY , we expect that $U(3)_{- \frac{5}{2}}$ decouples. 

There are also other special cases:

\vskip .1cm

- $\sigma^{N_c}_{N_c}= \sigma^{N_c-1}_{N_c-1} = \sigma^{N_c-2}_{N_c-2} = - m_1,  \sigma^{N_c-3}_{N_c-3} = 0$ with a gauge group
\be{3masses2333}
U(N_c - 3)_{k+4} \times U(3)_{k+\frac{3}{2}} 
\ee
- $\sigma^{N_c}_{N_c}= \sigma^{N_c-1}_{N_c-1} = \sigma^{N_c-2}_{N_c-2} = 0,  \sigma^{N_c-3}_{N_c-3} = - m_4$  with a gauge group
\be{3masses2344}
U(N_c - 1)_{k+4} \times U(1)_{k+\frac{7}{2}} 
\ee

- $\sigma^{N_c}_{N_c}= \sigma^{N_c-1}_{N_c-1} = - m_1, \sigma^{N_c-2}_{N_c-2} = \sigma^{N_c-3}_{N_c-3} = 0$ with a gauge group
\be{3masses2355}
U(N_c - 2)_{k+4} \times U(2)_{k+\frac{3}{2}} 
\ee

- $\sigma^{N_c}_{N_c}= - m_1, \sigma^{N_c-1}_{N_c-1} = \sigma^{N_c-2}_{N_c-2} = \sigma^{N_c-3}_{N_c-3} = 0$ with a gauge group
\be{3masses2366}
U(N_c - 1)_{k+4} \times U(1)_{k+\frac{3}{2}} 
\ee

- $\sigma^{N_c}_{N_c}= - m_1, \sigma^{N_c-1}_{N_c-1} = \sigma^{N_c-2}_{N_c-2} = \sigma^{N_c-3}_{N_c-3} =  - m_4$ with a gauge group
\be{3masses2377}
U(N_c - 2)_{k+4} \times U(1)_{k+\frac{3}{2}} \times   U(1)_{k+\frac{3}{2}}
\ee
 
$\bullet$ $m_1 = m_2 = m_3 < m_4$

Choose $\sigma^{N_c}_{N_c}= \sigma^{N_c-1}_{N_c-1} = \sigma^{N_c-2}_{N_c-2} = - m_1,  \sigma^{N_c-3}_{N_c-3} = - m_4$. For the first set 
of vevs of $\sigma$ one gets a $U(3)$ theory where 4 $\tilde{Q}$ with positive masses and one $Q$ with positive mass have been integrated out.
The effective level is then $k+\frac{5}{2}$. 

If $\sigma^{N_c-2}_{N_c-2} = - m_4$ we get $U(1)$ with 4  $\tilde{Q}$ with positive mass and three
 $Q$ with negative mass, the effective level is  $k+\frac{1}{2}$. The gauge group is 
\be{3massesin}
U(N_c - 4)_{k+4} \times U(3)_{k+\frac{5}{2}} \times U(1)_{k+\frac{1}{2}}
\ee
or in the $k = -4$ case
 \be{3massesm42}
U(N_c - 4)_{0} \times U(3)_{- \frac{3}{2}} \times U(1)_{-\frac{7}{2}}.
\ee
Among the cases with some zero values for the above $\sigma$ we mention the one with $\sigma^{N_c}_{N_c}= \sigma^{N_c-1}_{N_c-1} = \sigma^{N_c-2}_{N_c-2} = - m_1, 
\sigma^{N_c-2}_{N_c-2} = 0$ which provides the unbroken gauge group
\be{3massesm4445}
U(N_c - 3)_{0} \times U(3)_{- \frac{3}{2}} 
\ee
This is part of a candidate for Aharony duality because $U(3)_{- \frac{3}{2}}$ has a singlet dual. 

\subsection{Case (c): Two Masses are Equal}

We only consider the choice $m_1 = m_2 > m_3 > m_4$. Take the vacuum
\be{vac22}
\sigma^{N_c}_{N_c}= \sigma^{N_c-1}_{N_c-1}=- m_1,~\sigma^{N_c-2}_{N_c-2} = - m_3,~\sigma^{N_c-3}_{N_c-3} = - m_4.
\ee
corresponding to the gauge group
\be{twoequalmasses}
U(N_c - 4)_{k+4} \times U(2)_{k+1} \times U(1)_{k+\frac{5}{2}} \times  U(1)_{k+\frac{7}{2}}
\ee
For $k=-4$ this becomes
\be{twoequalmasses4}
U(N_c - 4)_{0} \times U(2)_{-3} \times U(1)_{-\frac{3}{2}} \times  U(1)_{-\frac{1}{2}}
\ee 
As in \cite{is}, $U(2)_{-3}$ with two massless flavors is an SCFT and should decouple in order to obtain an Aharony duality 
with singlets coming from $U(1)_{-\frac{1}{2}}$. 

One other important choice for two equal masses is $\sigma^{N_c}_{N_c}= \sigma^{N_c-1}_{N_c-1}= \sigma^{N_c-2}_{N_c-2} = 0,~\sigma^{N_c-3}_{N_c-3} = - m_4$ 
which provides an unbroken gauge group:
\be{twoequalmasses4444}
U(N_c - 1)_{0} \times  U(1)_{-\frac{1}{2}}
\ee 
containing a potential singlet coming from the dual of $U(1)_{-\frac{1}{2}}$.  
 
\subsection{Case (d): All Masses are Different}

We consider $m_1 > m_2 > m_3 > m_4$ and the following vacuum:
\be{vac4}
\sigma^{N_c-i}_{N_c-i}=- m_{i+1},~i=0,1,2,3.
\ee 
Each branch $\sigma^{N_c-i}_{N_c-i}=- m_{i+1}$ has 4 $\tilde{Q}$ with positive masses but the masses for $Q$ fields are positive or negative 
so the unbroken gauge group is
\be{diffmasses}
U(N_c - 4)_{k+4} \times U(1)_{k+\frac{1}{2}} \times U(1)_{k+\frac{5}{2}} \times  U(1)_{k+\frac{7}{2}}
\ee
becoming for $k=-4$:
\be{diff4}
U(N_c - 4)_{0} \times U(1)_{-\frac{7}{2}} \times U(1)_{-\frac{3}{2}} \times  U(1)_{-\frac{1}{2}}
\ee
One other important solution is $\sigma^{N_c}_{N_c}= \sigma^{N_c-1}_{N_c-1}= \sigma^{N_c-2}_{N_c-2} = 0,~\sigma^{N_c-3}_{N_c-3} = - m_4$ which 
contains a singlet coming from  $U(1)_{-\frac{1}{2}}$. The $ U(1)_{-\frac{7}{2}} \times U(1)_{-\frac{3}{2}}$ groups should 
decouple in order to preserve SUSY. 

\subsection{From Giveon-Kutasov to Aharony Duality for 4 Massive Flavors}

We consider a Giveon-Kutasov electric theory with gauge group $U(N_c)_{-4}$ with $N_f$ quark flavors with real masses for the $N_f,N_f-1, N_f-2,
N_f -3$ flavors. The Giveon-Kutasov magnetic dual is $U(N_f + 4 - N_c)_{-4}$. We integrate out the four massive quarks, choose the 
$\sigma = 0$ vacuum and obtain the Aharony electric theory $U(N_c)_{0}$ with $N_f - 4$ massless flavors. 

The modification of the Giveon-Kutasov magnetic picture depends on how we choose the masses for the 4 quarks, as discussed in the previous 
subsections:

$\bullet$ If all four masses are equal, we consider case (2a) and obtain the Aharony dual  with a gauge group     
$U(N_f - 4-N_c)_{0} \times U(4)_{-2} \times U(4)_{-2}$ w

$\bullet$ If only three  masses are equal, we  consider case (b) and obtain the Aharony dual with a gauge group which is either
\be{aea}
U(N_f - 4-N_c)_{0} \times   U(3)_{-\frac{5}{2}} \times U(3)_{-\frac{5}{2}} \times  U(1)_{-\frac{1}{2}} \times   U(1)_{-\frac{1}{2}}
\ee
or 
\be{beb}
U(N_f - 4-N_c)_{0} \times   U(3)_{-\frac{3}{2}} \times U(3)_{-\frac{3}{2}} \times  U(1)_{-\frac{7}{2}} \times   U(1)_{-\frac{7}{2}}
\ee
For the case (\ref{aea}) we get the two singlets $X_{\pm}$ from dualising $U(1)_{-\frac{1}{2}} \times   U(1)_{-\frac{1}{2}}$ and then decouple 
$U(3)_{-\frac{5}{2}} \times U(3)_{-\frac{5}{2}}$ by sending the masses to infinity. For the case (\ref{beb}) we get the singlets $X_{\pm}$ from 
dualising $U(3)_{-\frac{3}{2}} \times U(3)_{-\frac{3}{2}}$ and then decoupling $U(1)_{-\frac{7}{2}} \times   U(1)_{-\frac{7}{2}}$. 

Not all the gauge groups breaking give rise to Aharony duals. One example when we get Aharony dual is 
 $\sigma^{N_c}_{N_c}= \sigma^{N_c-1}_{N_c-1} = \sigma^{N_c-2}_{N_c-2} = 0,  \sigma^{N_c-2}_{N_c-2} = - m_4$ when the breaking is as in 
(\ref{3masses2344}) $U(N_c - 1)_{k+4} \times U(1)_{k+\frac{7}{2}}$. This becomes $U(N_c - 1)_{0} \times U(1)_{-\frac{1}{2}}$ when $k=-4$.
We recover Aharony  magnetic dual $U(N_f + 1-N_c)_{0} \times U(1)_{-\frac{1}{2}}$.

$\bullet$ If two masses are equal and we choose $m_1 = m_2 > m_3 > m_4$, the case (c) discussed above implies that the Aharony dual is
\be{cec}
U(N_f - 4-N_c)_{0} \times   U(2)_{-3}^2 \times U(1)_{-\frac{3}{2}}^2 \times  U(1)_{-\frac{1}{2}}^2 
\ee
with the two singlets $X_{\pm}$ being given by the duals of $U(1)_{-\frac{1}{2}}^2$. 

For $\sigma^{N_c}_{N_c}= \sigma^{N_c-1}_{N_c-1}= \sigma^{N_c-2}_{N_c-2} = 0,~\sigma^{N_c-3}_{N_c-3} = - m_4$ we have an unbroken electric gauge group
$U(N_c - 1)_{0} \times  U(1)_{-\frac{1}{2}}$  and unbroken magnetic gauge group  $U(N_f + 1 - N_c)_{0} \times  U(1)_{-\frac{1}{2}}$ which are indeed 
Aharony dual to each other.

$\bullet$ If all masses are different and we choose $m_1 > m_2 > m_3 > m_4$, the above case (d) implies that the Aharony magnetic dual is
\be{ded}
U(N_f - 4-N_c)_{0} \times   U(1)_{-\frac{7}{2}}^2 \times U(1)_{-\frac{3}{2}}^2 \times  U(1)_{-\frac{1}{2}}^2 
\ee
with the two singlets  $X_{\pm}$ being again the duals of $U(1)_{-\frac{1}{2}}^2$. The group $ U(1)_{-\frac{7}{2}}^2 \times U(1)_{-\frac{3}{2}}^2$ needs to decouple to preserve SUSY.
 
\section{Three Dimensional Theories with Adjoint Matter}

We consider the ${\cal{N}} = 4 \rightarrow {\cal{N}} = 2$ deformation for a general potential for the field $\Phi$ \cite{giku,kss}:
\be{def}
\int d^2 \theta \sum_{i=0}^{n} \frac{c_i}{n+1-i} \Phi^{n+1-i}
\ee
which implies that the superpotential has $n$ distinct minima $x=a_i$:
\be{adj1}
W'(x) = \sum_{j=0}^{n} c_i x^{n-i} = c_0 \prod_{i=1}^{n} (x - a_i)
\ee
The vacua are labeled by integers $(r_1,\cdots,r_n)$. When all the values of $a_i$ are distinct, the gauge group is Higgsed
\be{split}
U(N_c) \rightarrow U(r_1) \times U(r_2) \times \cdots \times U(r_n)
\ee
and we get $n$ decoupled copies of the  ${\cal{N}} = 2$ theories discussed in the previous section. 
   
If we add $N_f$ flavor fields, the full superpotential is
\be{phi1}
\int d^2 \theta {\sqrt 2}\Phi Q\Qt + \int d^2 \theta \sum_{i=0}^{n} \frac{c_i}{n+1-i} \Phi^{n+1-i}
\ee
The equation of motion for $\Phi$ is
\be{phi12}
\int d^2 \theta ({\sqrt 2} Q\Qt + \sum_{j=0}^{n} c_i \Phi^{n-i})
\ee 
and we take the simple solution $Q\Qt = W'(x) = 0$. 

The Giveon-Kutasov duality \cite{gk1} was generalized to the case with an adjoint field in \cite{niarchos}. 
The magnetic dual is an ${\cal{N}} = 2,~ U(n N_f + n|k| - N_c)$ theory with $N_f$ pairs of chiral multiplets $q \tilde{q}$, an
adjoint field $\Psi$ and $n$ magnetic mesons $M_i (i=1,\cdots,n)$.  On the other hand, the Aharony duality \cite{aha1} was generalized to the case with an adjoint field in \cite{park1}. The proposed magnetic dual is $U(n N_f - N_c)$ with $N_f$ pairs of chiral multiplets $q, \tilde{q}$, an
adjoint field $Y$, $n$ magnetic mesons $M_i (i=1,\cdots,n)$ and $2 n$ singlet
superfields $X_{\pm,1},\cdots,X_{\pm,n}$. The magnetic superpotential is 
\be{magpota}
W_m = Tr Y^{n+1} + \sum_{j=1}^{n-1} M_j \tilde{q} Y^{n-1-j} q + \sum_{i=1}^{n} (X_{+,i} \tilde{X}_{-,n+1-i} + 
X_{-,i} \tilde{X}_{+,n+1-i}). 
\ee
$X_{1,\pm}$ and $\tilde{X}_{1,\pm}$ are the monopoles of the electric and magnetic theory. The coupling between the monopoles is dictated by the requirement of having the right R symmetry charge together with the superconformal index matching  requirement \cite{park1}.

We now want to flow from a Giveon-Kutasov duality with adjoint matter to an Aharony duality with adjoint matter. In order to do so, we start by 
generalizing the results of \cite{is} to the case of theories with adjoint matter. 

\subsection{Symmetry Breaking for Theories with Adjoint Matter}

Let us consider the deformation ${\cal{N}} = 4 \rightarrow {\cal{N}} = 2$ for the 3 dimensional theory given by the superpotential (\ref{def}).
Each of the $U(r_i)$ groups in (\ref{split}) has $N_f$ fundamental flavor and we consider that each group has the same Chern-Simons level $k$. 

Fixing the expectation value for $\Phi$ does not imply that 
$\sigma$ is also fixed and, for each value $\Phi =  a_i$, we have a classical vacua encoded by 
$\sigma_i, Q^{N_f;i}, \tilde{Q}_{N_f;i}, i=1,\cdots,n$ 
obeying (\ref{lowest}) together with the D-term equation for each group $U(r_i)$. 

We now give a real mass to one or more flavors. If we give mass to only one flavor $Q^{N_f}$ and $\tilde{Q}_{N_f}$ , we have three cases:

$\bullet$ $\sigma_i,  Q^{N_f;i}= \tilde{Q}_{N_f;i}, i=1,\cdots,n$ are all zero. 

By integrating out the massive fundamental flavors, each $U(r_i)_k$ groups becomes $U(r_i)_{k+1}$ with $N_f - 1$ flavors. It is now easy to see what 
happens when the coefficients $c_i, i\ne 1$ in (\ref{def}) are zero, in which case the superpotential becomes $\Phi^{n+1}$. All $a_i$ 
become zero and the gauge group enlarges to $U(N_c)_{k+1}$ with $N_f - 1$ flavors.  

$\bullet$  $\sigma_{N_c;i}^{N_c;i} = m_i, i=1,\cdots,n$ as the only non-vanishing $\sigma$. Only one flavor is massive and the gauge group becomes
\be{breaks}
U(r_1-1)_{k+1} \times \cdots \times U(r_n-1)_{k+1}
\ee
with $N_f-1$ flavors each, together with $n$ decoupled $U(1)_{k+\frac{1}{2}}$ factors each having a single light field of charge $+1$ and 
a FI parameter $\xi = m_i(k+1)$.

If we now take  $c_i, i\ne 1$ in (\ref{def}) to zero, the  $U(r_i-1)_{k+1}$ factors in (\ref{breaks}) add up to 
$U(N_c-n)_{k+1}$ with $N_f - 1$ massless flavors together with a a product of $n$  $U(1)_{k+\frac{1}{2}}$ groups, each having a  FI parameter 
$\xi = m_i(k+1)$. 

$\bullet$  $\sigma_{N_c;i}^{N_i} = m_i, \sigma_{N_c-1;i}^{N_c-1;i} = - m_i, i=1,\cdots,n$ break the group to
\be{breaks1}
U(r_1-2)_{k+1} \times \cdots \times U(r_n-2)_{k+1}
\ee  
with $N_f-1$ massless flavors each together with a product of $n$ $U(1)_{k+\frac{1}{2}} \times U(1)_{k+\frac{1}{2}} $ groups, each $U(1)_{k+\frac{1}{2}}$
having a single charged field and a    FI parameter $\xi = \pm m_i(k+1)$.

If $c_i, i\ne 1$ are zero, all  $U(r_i-1)_{k+1}$ factors in (\ref{breaks1}) overlap and provide 
$U(N_c-2 n)_{k+1}$ with $N_f - 1$ massless flavors. The $n$  $U(1)_{k+\frac{1}{2}} \times U(1)_{k+\frac{1}{2}} $ factors are distinguished by the 
value of the FI parameter $\xi = \pm m_i(k+1)$. 

For $k=-1$, the above 3 cases become in the limit  $c_i, i\ne 1$ being zero:

1.  $U(N_c)_0$ with $N_f - 1$ flavors.

2.  $U(N_c-n)_{0}$ together with  $n$  $U(1)_{-\frac{1}{2}}$ groups.

3.  $U(N_c-2 n)_{0}$ together with  $n$  $U(1)_{-\frac{1}{2}} \times U(1)_{-\frac{1}{2}}$ groups.

We can apply the methods of the previous section to generalize this case in the presence of more massive fundamental flavors.

\subsection{Coupling between Monopoles}

Before moving to dualities, we need to address the question of the coupling between electric and magnetic monopoles in the potential dual theory. 
As in \cite{is}, the electric monopoles come from dualising the  $U(1)_{-\frac{1}{2}}$ into $X_{\pm}$ chiral superfields whereas the magnetic monopoles
are associated with $U(1)$ subgroups of the gauge groups. For the second case above, we can dualise the $U(1)_{-\frac{1}{2}}$ factors into $X_{+;i}, i=1,\cdots n$ chiral 
superfields, each one coupling to operators $\tilde{X}_{-;i}$ associated to $U(1)_i$ subgroups of $U(r_i-1)_{0}$ and a superpotential is generated as 
\be{pot12}
W = \sum_{i=1}^{n} X_{+;i} \tilde{X}_{-;i}
\ee 
For the third case above, we dualise one group of $U(1)_{-\frac{1}{2}}$ factors into $X_{+;i}, i=1,\cdots n$ and the other into $X_{-;i}, i=1,\cdots n$ which couple
to the magnetic monopoles as
\be{pot13}
W = \sum_{i=1}^{n} (X_{+;i} \tilde{X}_{-;i} + X_{-;i} \tilde{X}_{+;i})
\ee 

What happens if all $c_i, i\ne 1$ in (\ref{def}) are taken to zero and all the masses $m_i$ are equal? In this case all the monopoles  
$X_{\pm;i}, \tilde{X}_{\pm;i}, i = 1,\cdots,n$ for the cases 2 and 3 overlap and we need to reconsider their couplings. 
A brane configuration discussion cannot provide an answer regarding this coupling and we would need to lift our theory to an M-theory/F-theory
picture. Nevertheless, we can use the arguments of \cite{park1} concerning the matching of the chiral rings between magnetic and 
electric theories, together with considerations of the superconformal index for $U(N)$ theory with an adjoint.  

For a cubic superpotential in $\Phi$ ($n=2$) this would mean that  $X_{\pm;1}$ are identified with the electric monopoles and 
$\tilde{X}_{\pm;1}$ with the magnetic monopoles. From Table 2 in \cite{park1} we see that on one hand the state $X_{\pm;1}~\mbox{Tr} \Phi$ contributes to the index
and this is represented by a contribution  $X_{+;1}~\tilde{X}_{-;2} + X_{-;1}~\tilde{X}_{+;2}$ to the superpotential. On the other hand,
 $X_{\pm;2}$ and the $N_f \times N_f$ singlet $M_1$ would appear as singlets in 
the magnetic theory with no correspondent in the electric theory. They must be paired with monopole operators and disappear. The $M_1$ 
operator is canceled by $X_{+;2} \tilde{X}_{-;1} + X_{-;2} \tilde{X}_{+;1}$. The total contribution to the superpotential is then 
\be{n2}
 X_{+;1} \tilde{X}_{-;2} + X_{-;1} \tilde{X}_{+;2} +  X_{+;2} \tilde{X}_{-;1} + X_{-;2} \tilde{X}_{+;1}
\ee

For a quartic superpotential in $\Phi~(n=3)$, one has three electric monopoles and three magnetic monopoles. To see their coupling one can again consult Table 2 in \cite{park1} to see that  $X_{\pm;1} \tilde{X}_{\mp;3}$ and $X_{\pm;2} \tilde{X}_{\mp;2} $ contribute to the superconformal index whereas 
 $X_{\pm;3} \tilde{X}_{\mp;1}$ is needed to cancel unwanted singlets. For general $n$ we recover formula (\ref{magpota}).

\subsection{From Giveon-Kutasov duality to Aharony Duality for Adjoint Matter}

We start with a Giveon-Kutasov  electric theory $U(N_c)_{-1}$ with $N_f$ fundamental flavors and an adjoint field with a superpotential 
$\mbox{Tr}~\Phi^{n+1}$. By giving mass to one 
fundamental flavor and choosing $\sigma = 0$, we get the Aharony electric theory
 $U(N_c)_{0}$ with $N_f-1$ fundamental flavors and an adjoint field. 

To get to the Aharony magnetic dual, we start by considering a Giveon-Kutasov duality which takes the   $U(N_c)_{-1}$ with $N_f$ fundamental flavors and an adjoint field   to $U(n N_f + n - N_c)_{-1}$ with flavors and adjoint dual as in \cite{niarchos}. The Giveon-Kutasov magnetic theory is 
then deformed by a non-zero vev for $\sigma$ as in the third case above. The theory flows to one where the rank of the non-Abelian dual group 
decreases by $2 n$ and it becomes  $U(n N_f -  n - N_c)_{-1}$ with $N_f - 1$ fundamental flavors and an adjoint field.  We also have the 
$n$ products of $U(1)_{-\frac{1}{2}} \times U(1)_{-\frac{1}{2}}$ groups which are to be dualised into electric monopoles which couple to magnetic monopoles as above
\be{apsup}
X_{+,1} \tilde{X}_{-,n} + X_{-,1} \tilde{X}_{+,n} + \cdots +  X_{+,n} \tilde{X}_{-,1} + X_{-,n} \tilde{X}_{+,1}
\ee
We therefore found a pair of Aharony duals which are the same as  in \cite{park1}.  

\section*{Acknowledgments}

This work was supported in part by STFC.

\end{document}